\documentclass[osajnl,preprint,twocolumn,10pt]{revtex4-1}
\usepackage{amsmath}
\usepackage{graphicx}

\begin{document}
\title{Truly unentangled photon pairs without spectral filtering}

\author{Z. Vernon$^{1,2,*}$, M. Menotti$^2$, C.C. Tison$^{3,4,5}$, J.A. Steidle$^6$, M.L. Fanto$^{5,6}$, P.M. Thomas	$^6$, S.F. Preble$^6$, A.M. Smith$^5$, P.M. Alsing$^5$,  M. Liscidini$^2$, J.E. Sipe$^1$}
\affiliation{$^1$Department of Physics, University of Toronto, 60 St. George Street, Toronto, Ontario, Canada, M5S 1A7 \\ 
$^2$ Department of Physics, University of Pavia, Via Bassi 6, Pavia, Italy \\
$^3$ Florida Atlantic University, Boca Raton, Fl, 33431, USA \\
$^4$ Quanterion Solutions Incorporated, Utica, NY, 13502, USA \\
$^5$ Air Force Research Laboratory, Information Directorate, Rome, NY, 13441, USA \\
$^6$ Rochester Institute of Technology, Rochester, NY, 14623, USA}
\email{zachary.vernon@utoronto.ca}

\begin{abstract}
We demonstrate that an integrated silicon microring resonator is capable of efficiently producing photon pairs that are completely unentangled; such pairs are a key component of heralded single photon sources. A dual-channel interferometric coupling scheme can be used to independently tune the quality factors associated with the pump and signal and idler modes, yielding a biphoton wavefunction with Schmidt number arbitrarily close to unity. This will permit the generation of heralded single photon states with unit purity.
\end{abstract}

\maketitle
\section{Introduction}
Single photon sources are essential elements in a wide array of photonic quantum technologies. An ideal source would produce single photons deterministically, and promising strategies exist to achieve this \cite{Ding2016}. However, most  require cryogenic cooling, and are not easily integrated into a compact, CMOS-compatible chip-based device. A popular alternative approach is to design \emph{heralded} single photon sources \cite{Xiong2016}, in which photons are produced non-deterministically in pairs. One photon of such a pair heralds the existence of the other; post-selecting on detection of the herald photon then yields the desired counting statistics required from a single photon source. Since heralded single photon sources can exploit parametric fluorescence in passive nonlinear optical media -- either spontaneous four-wave mixing (SFWM) or spontaneous parametric down-conversion (SPDC) -- they can be designed to operate at room temperature, and are readily implemented in integrated devices. Many integrated photon pair sources for single photon heralding have recently been demonstrated on the rapidly developing platform of silicon nanophotonics \cite{Reimer2014,Steidle2016,Silverstone2015}.

A number of considerations are important when designing integrated heralded single photon sources: (i) the generation process should be efficient to permit operation at low pump powers; (ii) the simultaneous generation of multiple photon pairs must be suppressed to avoid heralding multi-photon states; (iii) vacuum amplitude must be avoided in the heralded state, ensuring a high heralding efficiency; (iv) the photons in each pair must be unentangled to yield heralded single photons in pure states, a property necessary to ensure high-visibility quantum interference \cite{Mosley2008,Cassemiro2010}. The first of these can be accomplished by using an optical microring resonator as the nonlinear medium, leading to dramatic enhancement in the generation efficiency of nonlinear processes \cite{Azzini2012,Vernon2015}. The second is achieved by pumping at an appropriately low power level to ensure the generation of only individual pairs. Appropriate mitigation of scattering losses through over-coupling the microresonator system can satisfy the third requirement \cite{Vernon2016,Steidle2016}.

Meeting the fourth requirement is a more subtle task. Path- and polarization-entanglement are naturally suppressed in integrated devices designed for single mode operation, but time-energy entanglement in photon pairs produced by parametric fluorescence is difficult to fully eliminate without compromising other aspects of the device. While spectral filtering of the generated photon pairs is often used, it necessarily reduces the number of photon pairs available, and degrades the symmetric heralding efficiency  of the source \cite{Mosley2008,Meyer-Scott2017}. Moreover, any additional optical element is a source of losses that can degrade the single-sided heralding efficiency of the device. A method for efficient and direct generation of truly unentangled pairs in an integrated device is the subject of this Letter. By leveraging recent advances in the design and fabrication of interferometric coupling schemes for SFWM in integrated silicon microring resonators \cite{Chen2007,Gentry2016}, we demonstrate a way to engineer fully separable biphoton wavefunctions using a device that is compatible with existing fabrication technology.

\section{Origin and impact of entanglement}
Microresonator-based photon pair sources driven by a pulsed pump prepare a quantum state of the form \cite{Helt2010}
$\vert \psi \rangle = c_0\vert \mathrm{vac}\rangle + \beta\vert \mathrm{II}\rangle + ...
$,
wherein $c_0$ is the vacuum amplitude, $\beta$ the two-photon amplitude (with $|\beta|^2$ the probability per pump pulse of generating a pair), and in which the ellipsis denotes higher-order terms that can safely be neglected at sufficiently low pump powers. Losses may lead to the contribution of single-photon amplitudes in the quantum state, but we assume these have been eliminated by suitably over-coupling the ring-channel system \cite{Vernon2016}. The two-photon state $\vert \mathrm{II}\rangle$ can be represented as
$\vert \mathrm{II}\rangle = \int d\omega_s d\omega_i \phi(\omega_s,\omega_i)a_S^\dagger(\omega_s)a_I^\dagger(\omega_i)\vert\mathrm{vac}\rangle,$
where $\phi(\omega_s,\omega_i)$ is the normalized biphoton wavefunction (BWF). The frequency variables $\omega_s$ and $\omega_i$ are understood to be respective offsets from the signal and idler reference frequencies $\omega_S$ and $\omega_I$, which correspond to two appropriately chosen resonances of the microring, and each integration is taken over a range large enough to capture the full extent of the corresponding resonance, but not so large as to overlap with any other resonances. The creation operators $a_S^\dagger$ and $a_I^\dagger$ are similarly defined only over the signal and idler resonance frequency ranges, respectively.

The detection of a signal photon heralds the existence of an idler photon, the quantum state of which is in general an incoherent mixture of single photon amplitudes within the idler resonance. The density matrix of the heralded idler photon, defined as $\rho_I=\mathrm{Tr}_S\lbrace \vert\mathrm{II}\rangle\langle\mathrm{II}\vert\rbrace$, is given by 
$\rho_I=\int d\omega d\omega' q_I(\omega,\omega') a_I^\dagger(\omega)\vert\mathrm{vac}\rangle\langle\mathrm{vac}\vert a_I(\omega')$,
where
$q_I(\omega,\omega') = \int d\omega'' \phi(\omega'',\omega)\phi^*(\omega'',\omega')$.
The purity $\gamma$ of $\rho_I$, defined by $\gamma=\mathrm{Tr}_I\lbrace \rho_I^2\rbrace$, is then simply
\begin{eqnarray}\label{purity_defn}
\gamma=\int d\omega d\omega' |q_I(\omega,\omega')|^2.
\end{eqnarray} 
A pure heralded idler photon (i.e. $\gamma=1$) is produced when the biphoton wavefunction $\phi(\omega_s,\omega_i)$ is fully separable, i.e., when $\phi$ can be expressed as
\begin{eqnarray}\label{eqn:separable_bwf}
\phi(\omega_s,\omega_i)=\phi_S(\omega_s)\phi_I(\omega_i)
\end{eqnarray}
for some normalized single-variable functions $\phi_S$ and $\phi_I$. The question then becomes: whence does the lack of separability in the biphoton wavefunction arise, and how can $\phi$ be engineered to satisfy (\ref{eqn:separable_bwf})?

For a microring resonator-based photon pair source point-coupled to a channel waveguide, the BWF can be expressed as  
\begin{eqnarray}\label{BWF_eqn}
\phi(\omega_s,\omega_i) \propto F_P(\omega_s + \omega_i) l_S(\omega_s)  l_I(\omega_i),
\end{eqnarray}
\cite{Vernon2015,Helt2010} (omitting proportionality factors unimportant for our present discussion), where
\begin{eqnarray}\label{pump_function}
F_P(\omega) = \int d\omega_p \alpha_P (\omega_p)\alpha_P(\omega-\omega_p)l_P(\omega_p)l_P(\omega-\omega_P),
\end{eqnarray}
for SFWM, and a similar expression holds for SPDC \cite{Helt2012}; since most demonstrations of photon pair generation in integrated microrings use $\chi_3$ materials, for the remainder of this Letter we focus on SFWM. In (\ref{BWF_eqn}) and (\ref{pump_function}) $l_J(\omega_j)=(-i\omega_j  +\omega_J/(2Q_J))^{-1}$, so that $|l_J(\omega_j)|^2$ is a Lorentzian function describing the resonance line at frequency $\omega_J$ with full width at half maximum (FWHM) $\omega_J/Q_J$; Here $Q_J$ is the corresponding loaded quality factor (which depends on both scattering losses in the ring and any coupling between the ring and bus waveguides). The function $F_P(\omega)$ depends on the pump spectral amplitude $\alpha_P(\omega_p)$, understood to describe a pump pulse with spectrum centred on $\omega_P$ so that $\alpha_P(\omega_p)$ is peaked around $\omega_p=0$. The pump centre frequency $\omega_P$ is also chosen to match one of the ring resonances, and is assumed to satisfy $2\omega_P=\omega_S+\omega_I$ for SFWM, or $\omega_P=\omega_S+\omega_I$ for SPDC.

The function $F_P(\omega)$ can be understood as the coherent sum of all energy-conserving amplitudes for two pump photons \emph{in the ring} to be converted to a signal and idler photon pair with total energy $\hbar\omega_S+\hbar\omega_I +\hbar\omega$. In the limit of a cw pump  $F_P(\omega)$ is narrowly peaked about $\omega=0$, so that $|F_P(\omega_s+\omega_i)|^2$ has support only along the line $\omega_s = -\omega_i$, giving rise to a BWF that is highly anti-correlated in energy \cite{Vernon2015,Helt2010}. This extreme lack of separability of the BWF can be mitigated by broadening $F_P(\omega)$ by an amount sufficient to make $|F_P(\omega_s+\omega_i)|^2$ nearly uniform over the domain defined by the Lorentzian factors in (\ref{BWF_eqn}). This broadening can be accomplished by using a sufficiently spectrally broad pump, relaxing the strict correlation of the generated photon energies to the central frequency of the pump pulse. By using pump photons with a large spread in energy, photon pairs can be generated that are not strictly anti-correlated in their offsets from their respective resonances.

Broadening the pump pulse spectrum in this manner to fully excite the pump resonance drastically reduces the degree of correlation in the BWF. However the filtering of the pump light as it enters the resonator (represented by the Lorentzian factors $l_P(\omega_P)l_P(\omega-\omega_P)$ in the integrand of (\ref{pump_function})) means that the spectrum of pump photons available for conversion cannot be arbitrarily increased by broadening the injected pump, and rather is fundamentally limited by the linewidth of the pump resonance. 

Therefore to achieve a truly uncorrelated BWF the pump resonance linewidth must be significantly broader than those of the signal and idler; equivalently, \textbf{the pump resonance quality factor must be much smaller than those of the signal and idler resonances.} Most examples in the literature of microring-based SFWM have used devices with nearly equal resonance linewidths for the pump, signal and idler, yielding BWFs that display residual correlations even when pumped by a pulse with an arbitrarily broad spectrum. This is illustrated in Fig. \ref{jsi_comparison}a, in which the joint spectral intensity (JSI) $\Phi(\omega_s,\omega_i)=|\phi(\omega_s,\omega_i)|^2$ for such a system is displayed alongside the Schmidt number $K$ for the corresponding BWF that quantifies its lack of separability \cite{Law2004,Quesada2015}; For a microring with equal quality factors, pumped with a Gaussian pulse, ignoring any dispersion we find that the lower bound is $K=1.08$ (in line with previous results \cite{Helt2010,Silverstone2015,Grassani2016}), significantly higher than the ideal $K=1$ enjoyed by a truly separable state; the corresponding upper bound on the purity of the heralded idler state is $\gamma=93\%$, which in this case can be calculated as $\gamma=1/K$.

\section{Interferometric coupling}
\begin{figure}
\centering
\includegraphics[width=0.9\linewidth]{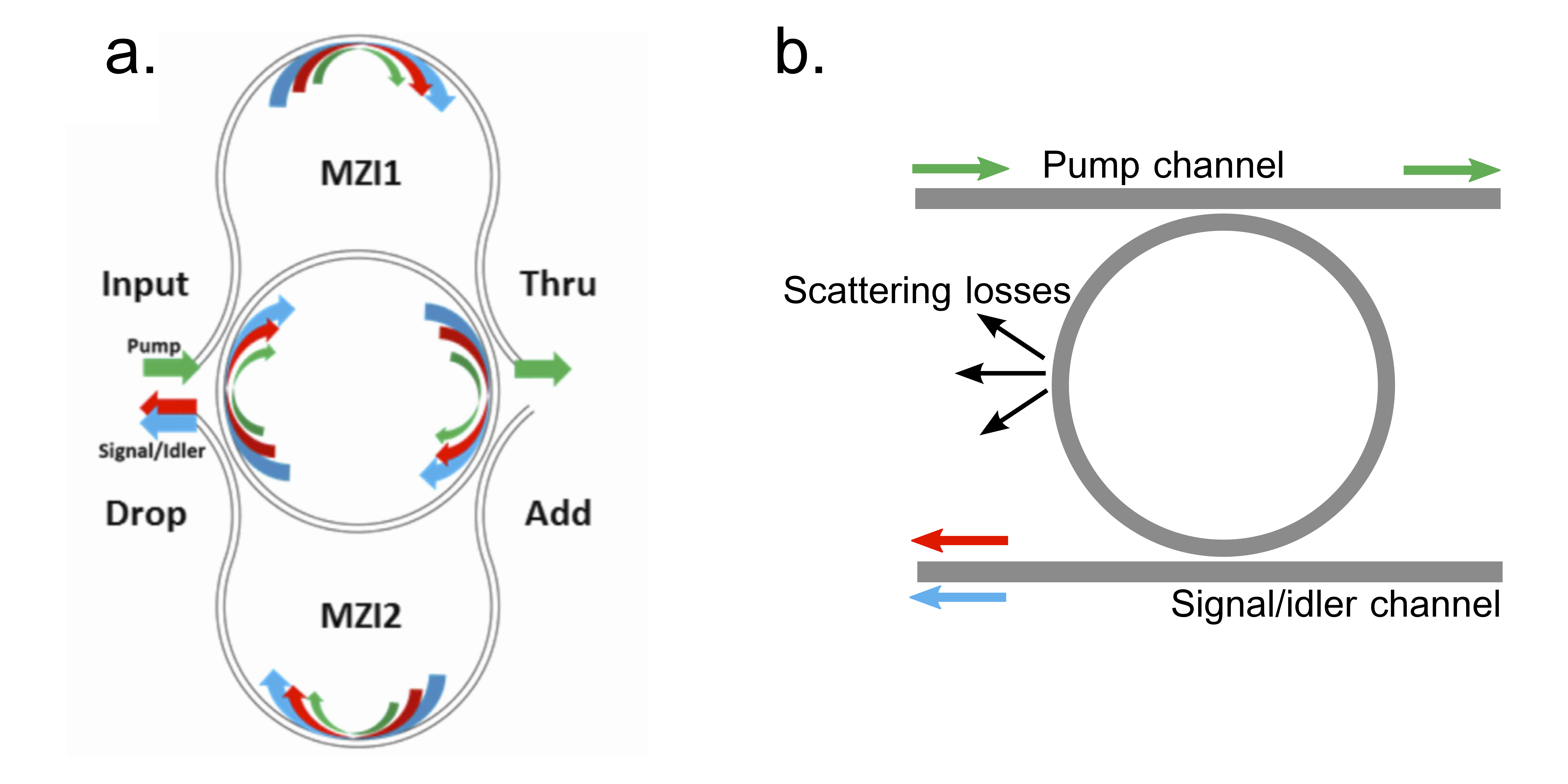}
\caption{(a) Design of a microring device with an interferometric coupling
scheme (MZI1 \& MZI2), permitting independent tuning of the pump, signal, and
idler resonances. (b) Idealized point-coupling model of dual-MZI device, with separate channels for the pump and signal/idler introduced, each with independent couplings to the ring.}\label{rit_schematic}
\end{figure}
To fully eliminate spectral correlations in microresonator-based photon pair sources, one must be able to tune the quality factors of the pump and signal/idler resonances independently. In conventional devices this is not possible, and usually the resonator-channel couplings and intraresonator losses do not vary significantly over the working frequency ranges. 

This limitation can be remedied by changing the coupling schemes between the channel and resonator \cite{Tison2017}. Recently, Gentry \emph{et al.} \cite{Gentry2016} showed that two coupling points between the bus waveguide and the ring provide a precise control over the individual quality factors of the ring resonances, and therefore also over the quantum properties of the generated light. The presence of two coupling points establishes a Mach-Zehnder interferometer (MZI); it allows one to control the interference between the light in the ring and that in the bus waveguide by using microheaters mounted on the device. Interferometric coupling of this type overcomes the usual limits on the separability of the biphoton state that plague microresonator systems lacking independent control over their different resonances. 

Here we consider the system shown in Fig. \ref{rit_schematic}a, which consists of a ring and two channels, and focus on leveraging its MZI-based coupling scheme to generate spectrally pure heralded single photons via the strategy discussed in Sec. 2. By appropriately tuning the MZIs, the system can be configured such that the top channel only couples to the pump mode in the resonator, while the bottom channel only couples to the signal and idler modes. Increasing the coupling strength of the top channel then decreases the quality factor of the pump resonance without significantly affecting the properties of the signal and idler resonances, allowing pump light with a much larger bandwidth to couple into the ring.

\begin{figure}[htb!]
\centering
\includegraphics[width=0.8\linewidth]{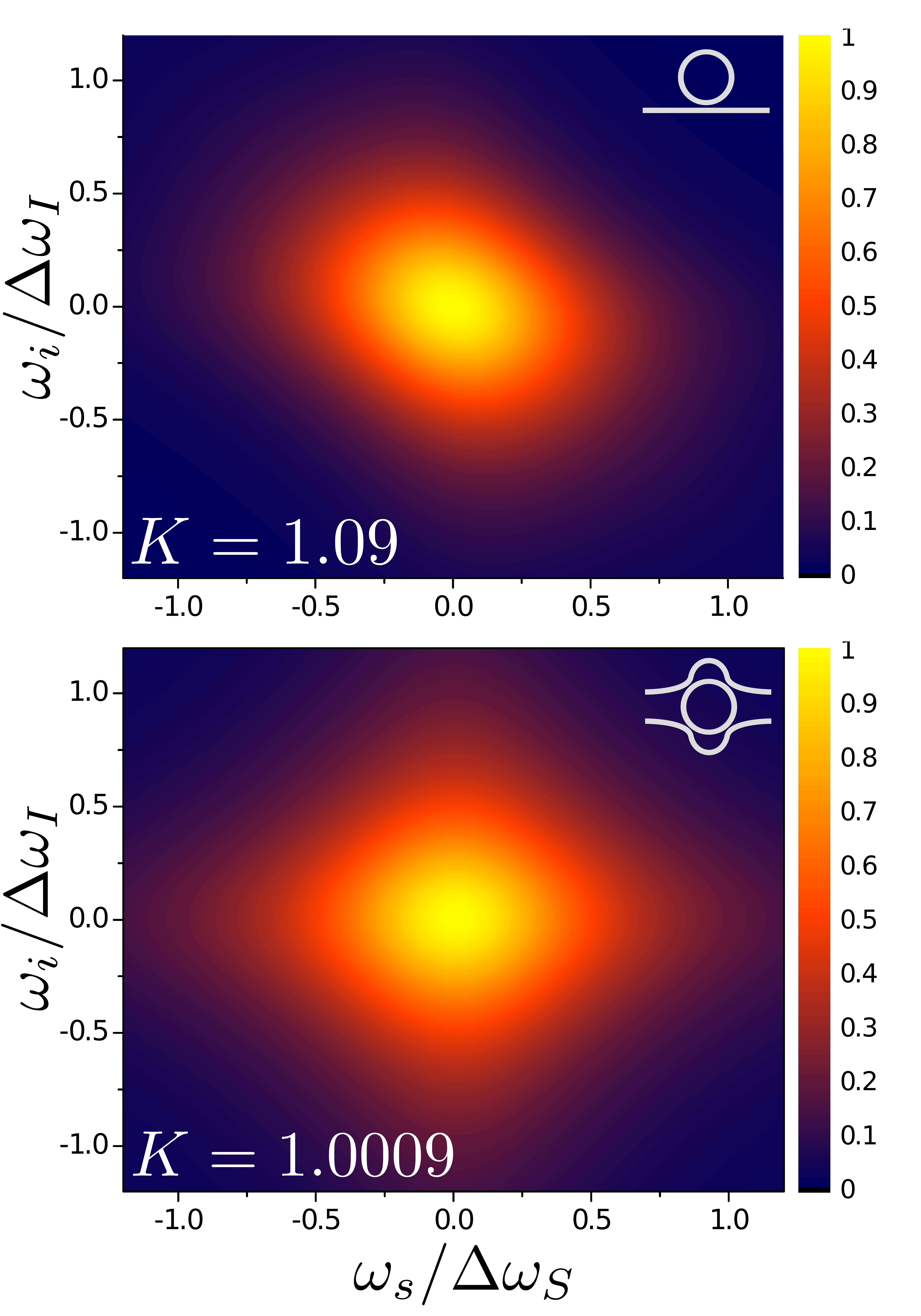}
\caption{(a) JSI for conventional single-channel system with equal quality factors, exhibiting residual anti-correlations that leads to a Schmidt number $K=1.09$ for the corresponding BWF (giving heralded state purity $\gamma=0.93$). (b) JSI for system with pump quality factor 6.6 times lower than the smallest of the signal/idler quality factors; the energy correlation has been almost completely eliminated, yielding $K=1.0009$ ($\gamma=0.999$). Both JSIs assume a pump pulse sufficiently broad to excite the entire pump resonance. The frequency axes -- offsets from the reference frequencyes $\omega_J$ -- are plotted in units of the respective signal/idler linewidths $\Delta\omega_J=\omega_J/Q_J$.}\label{jsi_comparison}
\end{figure}

\begin{figure}[htb!]
\centering
\includegraphics[trim={0cm 3.0cm 0cm 3cm},clip,width=1.0\linewidth]{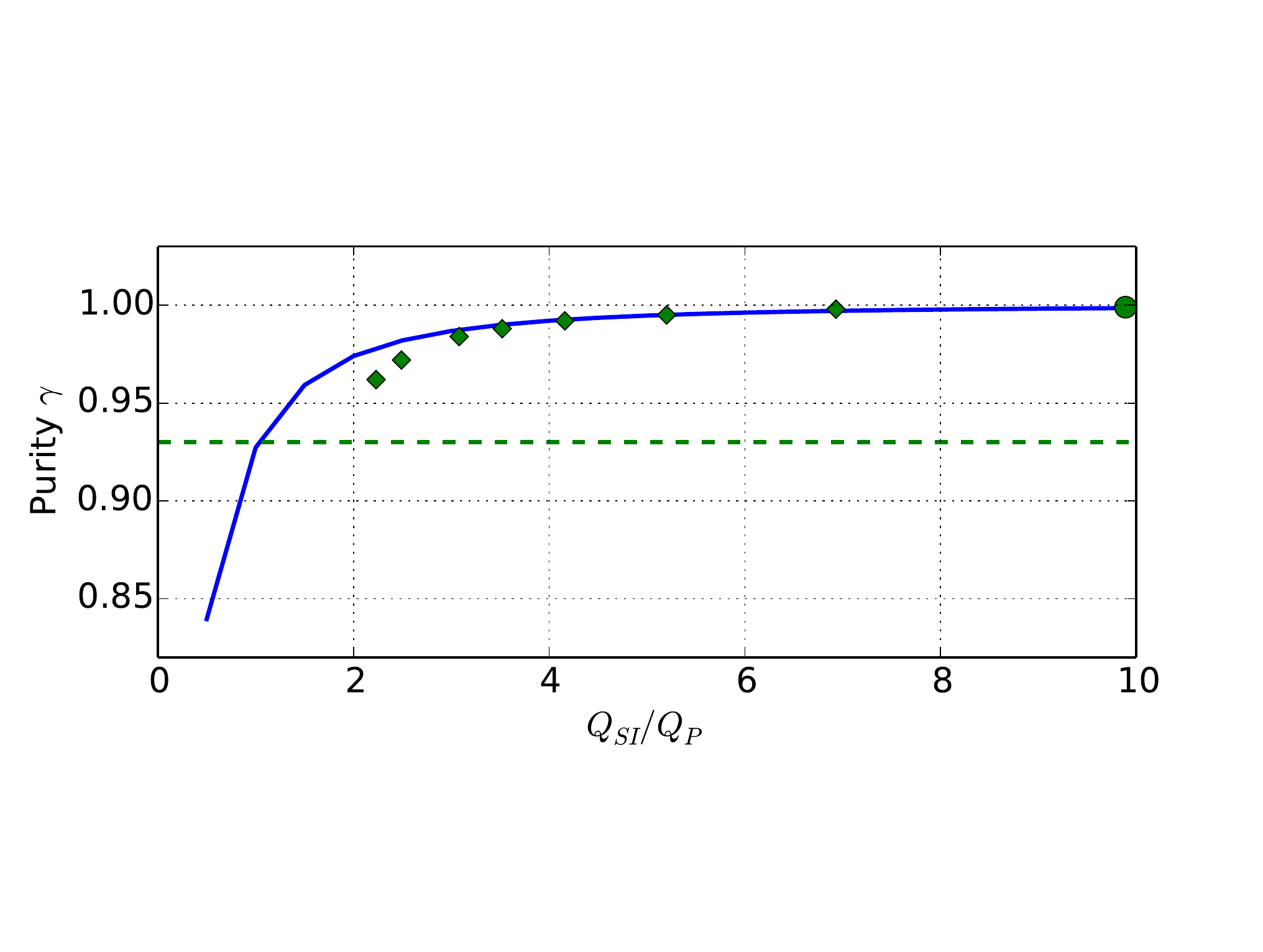}
\caption{Maximum heralded state purity as a function of ratio $Q_{SI}/Q_P$, with $Q_{SI}=\min(Q_S,Q_I)$. The solid line is calculated using an effective point-coupling model for the BWF; green markers indicate the results of full numerical simulations for the complete device, with the circular marker corresponding to the BWF of Fig. \ref{jsi_comparison}b. Each point assumes a pump pulse sufficiently broad to excite the full pump resonance. The dashed line represents the upper bound of $93\%$ purity for a system with equal quality factor, which is clearly exceeded with the proposed strategy.}\label{purity_plot}
\end{figure}

 In Fig. \ref{jsi_comparison} we compare the JSI distributions for a conventional device and one with an interferometric coupling, demonstrating the improvement in separability. These distributions -- and the corresponding complete BWFs -- were calculated using full numerical simulations of the asymptotic fields for the devices \cite{Liscidini2012}, and are in agreement with the predictions of the analytic expression (\ref{BWF_eqn}), verifying the applicability of an effective point-coupling model with independently set couplings for the pump and signal/idler (represented in Fig. \ref{rit_schematic}b) to the interferometrically coupled device (\ref{rit_schematic}a). The phase of the BWF corresponding to the JSI in Fig. \ref{jsi_comparison}(b) was also found to be uncorrelated, proving the lack of entanglement in the corresponding biphoton state.

To further examine the improvement offered by interferometric coupling, it is helpful to study the dependence of the heralded state purity $\gamma$ (\ref{purity_defn}) as a function of the ratio $Q_{SI}/Q_P$, where $Q_{SI}$ is the minimum of the signal and idler mode quality factors \cite{Q_footnote}. This dependence is illustrated in Fig. \ref{purity_plot}, showing how the device can be used to exceed the $93\%$ upper bound on the purity for a conventional system with equal quality factors. The solid curve corresponds to BWFs calculated using the simplified model (\ref{BWF_eqn}) which assumes point-coupling (Fig. \ref{rit_schematic}b), with the green markers indicating the results of full numerical simulations of the device in Fig. \ref{rit_schematic}a. 

It is important to ensure that the proposed strategy to eliminate time-energy entanglement does not seriously compromise any features of an ideal heralded single photon source (see Sec. 1). This can be addressed by comparing a conventional, single-channel microring system with equal quality factors for the three modes, all strongly over-coupled to ensure high heralding efficiency, to the new device with independently tunable quality factors.  It is clear that simultaneous generation of multiple photon pairs is still suppressible in the new device by using sufficiently low pump powers, and that collection of both pair photons can still be achieved by over-coupling the signal/idler channel with respect to scattering losses. However, one might initially suspect that the generation efficiency of the new device would be significantly worse than the conventional device due to the lowering of the pump quality factor to ensure high purity of the heralded photon. 

In fact the degradation of the generation efficiency is not expected to be especially serious. A full derivation \cite{Vernon2015,Vernon2016} shows that the rate of generated photon pairs available in the signal/idler channel for the idealized device shown in Fig. \ref{rit_schematic}b can be expressed as $J_\mathrm{heralds}=4\Lambda^2 q_Sq_Iq_P^2 f_P \mathcal{E}_\mathrm{P}^2 \tau^2 I/(\pi^3 (\hbar\omega_P)^2)$, where $\Lambda$ is a constant related to the nonlinear response of the microring \cite{Vernon2015b}, $q_J=Q_J/Q_J^\mathrm{ext}$ is the ratio for resonance $J$ between the loaded quality factor $Q_J$, which takes into account both scattering losses and the resonator-channel coupling, and the extrinsic quality factor $Q_J^\mathrm{ext}$ related only to the ring-channel coupling, $f_P$ is the repetition rate of the pump laser, $\mathcal{E}_\mathrm{P}$ the pump pulse energy, and $\tau$ the pump pulse duration. The dimensionless factor $I$ is related to the overlap between the intra-resonator pump amplitudes and the Lorentzian functions describing the signal and idler resonances. This overlap improves as $Q_{S,I}/Q_P$ increases; for a Gaussian pump pulse with duration $\tau$ having temporal intensity profile proportional to $e^{-t^2/\tau^2}$, we have
\begin{eqnarray}
\lefteqn{I=}\\
& &\int du_s du_i\frac{\bigg\vert \int du_p \frac{e^{-(T_Pu_p)^2/2}e^{-(T_Su_s+T_Iu_i-T_Pu_p)^2/2}}{(-iu_p+1)(-i(r_Su_s+r_Iu_i-u_p)+1)}\bigg\vert^2}{(u_s^2+1)(u_i^2+i)}, \nonumber
\end{eqnarray}
where $T_J=\tau/\tau^\mathrm{dwell}_J$ is the pulse duration in units of the photon dwelling time $\tau^\mathrm{dwell}_J=2Q_J/\omega_J$ for resonance $J$, and $r_J=Q_P/Q_J$. To fully excite the pump resonance we must choose a pulse with sufficiently short duration that $T_P=1$, so that $J_\mathrm{heralds}\propto q_P^2Q_P^2\mathcal{E}_P^2I$; or, in terms of the peak pump power $P_\mathrm{peak}\propto \mathcal{E}_P/\tau$, $J_\mathrm{heralds}\propto q_P^2Q_P^4 P_\mathrm{peak}^2 I$. As the channel-ring coupling for the pump is increased (so that $Q_P$ is decreased) and the pulse duration is lowered to fully excite the pump resonance and keep $T_P=1$: (i) $q_P$ increases toward unity; (ii) $Q_P$ decreases; (iii) $I$ increases asymptotically to $I_\mathrm{asy}\approx 14.3$ (from $I\approx 5.0$ when all quality factors are equal). A system with initially equal quality factors, strongly over-coupled for all resonances, with $Q_P$ subsequently altered to obtain $Q_{S,I}/Q_P=10$ requires a peak power about 61 times larger to maintain the original pair generation rate. Yet this corresponds to merely a $6.1$x increase in pulse energy, since the pulse duration has been compressed by a factor of $10$. Though this discussion of generation efficiency strictly  applies only to an idealized model of the full device, and neglects the possibility of photon pair generation in the outer arms of the interferometric couplers, it serves to demonstrate that only modest increases in energy requirements are anticipated; these are not expected to present a challenge in experiments. A study of the generation efficiency in the full device shown in Fig. \ref{rit_schematic}a is left for future communications.

\section{Conclusion}
We have shown that photon pairs with fully separable biphoton wavefunctions can be generated efficiently in an integrated device using a  microring resonator in an interferometrically-coupled dual channel configuration. This permits the generation of heralded single photon states with purity arbitrarily close to $100\%$, overcoming the upper bound of $93\%$ for a Gaussian pump pulse in conventional microresonator systems without the use of spectral filtering or sophisticated phase matching techniques \cite{Chen2017}. The strategy presented does not seriously compromise the heralding efficiency and generation efficiency. We expect this to have an impact on the design and implementation of integrated quantum photonic technologies that require reliable on-chip sources of indistinguishable single photons.

JES and ML acknowledge European Cooperation in Science and Technology (COST) (MP 1403). JAS, ZW, NAM,  and SFP acknowledge partial support under NSF ECCS-1408429.
CCT, MLF, AMS, and PMA acknowledge support from OSD ARAP QSEP program. JAS, PMT, and SFP acknowledge support under AFRL grant FA8750-16-2-0140. Any opinions, findings and conclusions or recommendations expressed in this material are those of the author(s) and do not necessarily reflect the views of AFRL.
\bibliography{jsi_letter}

\begin{thebibliography}{10}
\bibitem{Q_footnote}
In tuning the device it is not always possible to maintain equal signal and
  idler quality factors. However, this does not affect our results, as the
  absolute scale of the signal and idler linewidths has no impact on the
  separability of the biphoton wavefunction; we merely require the pump quality
  factor $Q_P$ to be smaller than the minimum of $Q_S$ and $Q_I$.
\end{thebibliography}

\end{document}